\documentclass[11pt,a4paper]{article}
\pdfoutput=1 
\usepackage{amsmath}
\usepackage{amssymb}
\usepackage{amsthm}
\usepackage{enumerate}
\usepackage{multirow}
\usepackage{mathtools}
\usepackage{jinstpub}
\usepackage{lineno}
\arxivnumber{1703.07214}
\usepackage{comment}
\usepackage[separate-uncertainty,retain-explicit-plus,per-mode = symbol]{siunitx}
\usepackage{url}
\newcommand{\pc}{\mbox{PC}}
\newcommand{\tmb}{\mbox{TMB}}
\newcommand{\ppo}{\mbox{PPO}}
\newcommand{\bet}{\mbox{$\beta$}}
\newcommand{\alp}{\mbox{$\alpha$}}
\newcommand{\gr}{\mbox{$\gamma$-ray}}
\newcommand{\grs}{\mbox{$\gamma$-rays}}
\newcommand{\E}[1]{\mbox{$\times10^{#1}$}}

\newcommand{\lith}{\mbox{$^{7}$Li}}

\newcommand{\ber}{\mbox{$^{7}$Be}}

\newcommand{\borten}{\mbox{$^{10}$B}}

\newcommand{\amer}{\mbox{$^{241}$Am}}

\DeclareSIUnit\c{$c$}
\DeclareSIUnit\week{w}
\DeclareSIUnit\year{yr}
\DeclareSIUnit\standard{std}
\DeclareSIUnit\str{sr}
\DeclareSIUnit\pe{PE}
\DeclareSIUnit\spe{SPE}
\DeclareSIUnit\ev{events}
\DeclareSIUnit\bin{(5-PE bin)}
\DeclareSIUnit\hit{hits}
\DeclareSIUnit\sgm{$\sigma$}
\DeclareSIUnit\rms{RMS}
\DeclareSIUnit\keVr{keV$_{\rm r}$}
\DeclareSIUnit\keVee{keV$_{\rm $e$e}$}
\DeclareSIUnit\ph{photons}
\DeclareSIUnit\neu{neutrons}
\DeclareSIUnit\inch{''}
\DeclareSIUnit\bit{bit}
\DeclareSIUnit\sample{samples}
\DeclareSIUnit\barn{b}

\newcommand{\pe}{\mbox{PE}}
\newcommand{\spe}{SPE}

\newcommand{\keVee}{\,keV$_{\text{\rm 	e}e}$}
\newcommand{\keVr}{\,keV$_{\rm r}$}

\newcommand{\geant}{GEANT4}

\newcommand{\scene}{\mbox{SCENE}}
\newcommand{\dsf}{\mbox{DarkSide-50}}

\newcommand{\snop}{\mbox{SNO+}}
\newcommand{\kamland}{\mbox{KamLAND}}
\newcommand{\pmt}{\mbox{PMT}}

\newcommand{\chisqndf}{\mbox{$\chi^{2}$/NDF}}

\title{Quenching Measurements and Modeling of a Boron-Loaded Organic Liquid Scintillator}
\author[a,b,1]{S.~Westerdale
\note{Corresponding author.}}
\emailAdd{shawest@physics.carleton.ca}
\author[a,c]{J.~Xu}
\author[a]{E.~Shields}
\author[a,d]{F.~Froborg}
\author[a]{F.~Calaprice}
\author[e,f]{T.~Alexander}
\author[g]{A.~Aprahamian}
\author[e,f]{H.O.~Back}
\author[g]{C.~Casarella} 
\author[g,h]{X.~Fang} 
\author[g,i]{Y.K.~Gupta}
\author[g]{E.~Lamere}
\author[g]{Q.~Liu} 
\author[g,j]{S.~Lyons}
\author[g,j]{M.~Smith} 
\author[g]{W.~Tan}

\affiliation[a]{Department of Physics, Princeton University, Princeton, NJ 08544, USA}
\affiliation[b]{Department of Physics, Carleton University, Ottawa, ON K1S 5B6, Canada}
\affiliation[c]{Lawrence Livermore National Laboratory, 7000 East avenue, Livermore, CA 94550}
\affiliation[d]{Imperial College London, Department of High Energy Physics, Blackett Laboratory, London SW7 2BZ, United Kingdom}
\affiliation[e]{Fermi National Accelerator Laboratory, Batavia, IL 60510, USA}
\affiliation[f]{Pacific Northwest National Laboratory, Richland, WA 99352, USA}
\affiliation[g]{Department of Physics, University of Notre Dame, Notre Dame, IN 46556, USA}
\affiliation[h]{Sino-French Institute of Nuclear Engineering and Technology, Sun Yat-Sen University, Zhuhai 519082, PR China}
\affiliation[i]{Nuclear Physics Division, BARC, Mumbai 400085, India}
\affiliation[j]{National Superconducting Cyclotron Laboratory, Michigan State University, East Lansing, MI 48824, USA}

\date{\today}

\abstract{
Organic liquid scintillators are used in a wide variety of applications in experimental nuclear and particle physics. Boron-loaded scintillators are particularly useful for detecting neutron captures, due to the high thermal neutron capture cross section of $^{10}$B. These scintillators are commonly used in neutron detectors, including the DarkSide-50 neutron veto, where the neutron may produce a signal when it scatters off protons in the scintillator or when it captures on $^{10}$B. Reconstructing the energy of these recoils is complicated by scintillation quenching. Understanding how nuclear recoils are quenched in these scintillators is an important and difficult problem.
In this article, we present a set of measurements of neutron-induced proton recoils in a boron-loaded organic liquid scintillator at recoil energies ranging from 57--467 keV, and we compare these measurements to predictions from different quenching models. We find that a modified Birks' model whose denominator is quadratic in $dE/dx$ best describes the measurements, with $\chi^2$/NDF$=1.6$. This result will help model nuclear recoil scintillation in similar detectors and can be used to improve their neutron tagging efficiency.
}


\begin{document}
\maketitle
\flushbottom


\section{Introduction}
Organic liquid scintillators are used in many different applications to measure the energy depositions by charged particles. These scintillators produce a number of photons approximately linearly proportional to the energy deposited in them by high energy electron recoils. This property makes these liquids useful for measuring electron recoil energy spectra such as those from \bet\ and \gr\ radiation. As a result, organic liquid scintillators are used in a wide variety of experiments, including Borexino~\cite{Alimonti:1998kw}, \snop~\cite{okeeffe_scintillation_2011}, SABRE~\cite{tomei_sabre}, and \kamland~\cite{eguchi_first_2003}. These scintillators are also often used to detect nuclear recoils; however, the high Linear Energy Transfer (LET) of these nuclei suppresses their scintillation yield nonlinearly and makes their energies harder to reconstruct.

Boron-loaded organic liquid scintillators are particularly useful for detecting neutrons. Approximately 20\% of natural boron is \borten, which has a high thermal neutron capture cross section of 3838\,b~\cite{Wright:2011ig}, and captures neutrons through the reaction

\begin{equation*}
^{10}\text{B}+n\rightarrow\begin{cases}^7\text{Li (1015)}+\alpha \text{ (1775)} & \mbox{(6.4\%)}\\
^7\text{Li}^* + \alpha \text{ (1471)}, & \mbox{(93.6\%)}\\ \qquad ^7\text{Li}^*\rightarrow ^7\text{Li (839)} + \gamma \text{ (478)} \end{cases}
\label{eq:bortencapture}
\end{equation*}
where the energy of each product is given in keV in parentheses. 

Boron-loaded scintillators are discussed in detail in~\cite{greenwood_improved_2008}. As discussed in~\cite{chou_integral_1993,pino_detecting_2014}, these scintillators are used in a wide range of applications that rely on neutron detection, including the \dsf~\cite{Agnes:2015gu} neutron veto~\cite{westerdale_prototype_2016,agnes_veto_2016}, anti-neutrino detection experiments~\cite{wang_feasibility_1999}, parity violation studies~\cite{yen_high-rate_2000}, and neutron monitoring applications~\cite{rasolonjatovo_development_2002,swiderski_boron-10_2008}. Since neutron detection relies on detecting nuclear recoils---either from the neutron scattering on nuclei in the scintillator or from the capture products on \borten---reconstructing the energy of events seen in the detector requires an understanding of the nuclear recoil scintillation.


Nuclear recoils tend to be heavily suppressed in organic liquid scintillators, and they tend to exhibit a high degree of nonlinearity in the scintillation response. Since detectors are typically calibrated to electron recoils at a small set of recoil energies, modeling this nonlinearity relative to the response of electron recoils is key to reconstructing the energy of nuclear recoils. This scintillation quenching effect was first described by Birks~\cite{birks_theory_1965} as a result of inter-molecular interactions between species produced at high ionization densities allowing excitation energy to non-radiatively dissipate. The result of these interactions is that energy depositions that occur with higher LET are more heavily suppressed than those with lower LET. However, observations have shown that this model is incomplete and does not accurately describe quenching at low proton recoil energies~\cite{robertson_quenching_2013}. 
Furthermore, since this mechanism depends on the species that are produced by ionizing radiation, introducing new compounds to a scintillator cocktail, such as a boron loading agent, may influence the quenching mechanism. 

This nonlinearity has been previously explored in commercial organic liquid scitillators in~\cite{pozzi_analysis_2004,enqvist_neutron_2013,iwanowska_time_flight_2015}, which fit empirical quadratic and exponential functions to the detector response relating the deposited energy to the observed electron equivalent energy. However, these studies all focused on recoil energies above a few hundred keV, with few to no measurements near the Bragg peak, which tends to be $\sim100$\,keV. The lowest energy recoils reported in~\cite{enqvist_neutron_2013} show that these models start to diverge at the lowest energies measured, indicating that a more complete model accounting for the stopping power may be needed to describe the quenching of lower energy proton recoils.

Many of the optical effects of boron-loading via the addition of trimethyl borate (\tmb) have been studied in~\cite{westerdale_prototype_2016,westerdale_thesis}. In particular, it has been shown that diluting a pure pseudocumene (\pc) and 2,5-Diphenyloxazole (\ppo) scintillator with equal parts \tmb\ decreases the light yield by $\sim15\%$ and that a high light yield can be achieved using a high concentration of \tmb. These properties make \tmb\ an effective boron-loading agent. 

Boron-loaded liquid scintillator detectors that primarily rely on detecting neutron captures may suffer from poor timing resolution (due to the thermal neutron capture time being on the scale of 2--20\,$\mu$s for typical \tmb\ concentrations~\cite{westerdale_thesis}), lack of neutron energy resolution, and poor tagging efficiency when the neutron escapes before capturing or when the capture signals fail to be detected. By also tagging neutrons when they scatter in the scintillator, these factors can be improved. Such improvements rely upon an accurate understanding of the signal that neutrons produce when they do so. 

In an organic scintillator, neutrons lose most of their energy to recoils off of hydrogen nuclei. We therefore present measurements of proton recoil quenching factors in a boron-loaded scintillator comprised of equal volumes of \pc\ and \tmb, with 3\,g/L of a \ppo\ wavelength shifter. 

We explore three different models that have been suggested for describing the quenching of organic liquid scintillators. We compare our measurements to these models in order to determine which best describes our observations. In particular, we study the models described in~\cite{birks_scintillations_1951,hong_scintillation_2002,craun_analysis_1970}. The only property of the recoiling nucleus that these models depend on is the stopping power. Since Birks' model has been shown to disagree with data at high stopping powers~\cite{robertson_quenching_2013}, an improvement in our ability to model low energy proton recoils may translate to a better ability to model quenching of heavier nuclei, such as those produced by neutron captures. However, since this study focuses on proton recoils, we do not test the models' ability to extrapolate to heavier nuclei here.

This experiment was inspired by the success of the \scene\ experiment, which measured the effects of nuclear recoil quenching in liquid argon at low recoil energies ($\gtrsim10$\,keV)~\cite{Cao:2015ks,Alexander:2013ke}. Data were taken using the same experimental setup as described in~\cite{xu_scintillation_2015}, which measured nuclear recoil quenching factors in a NaI(Tl) crystal.

\section{Experimental setup}\label{sec:setup}
\subsection{Overview}
\begin{figure}[htb]
 \centering
 \includegraphics[width=0.8\linewidth]{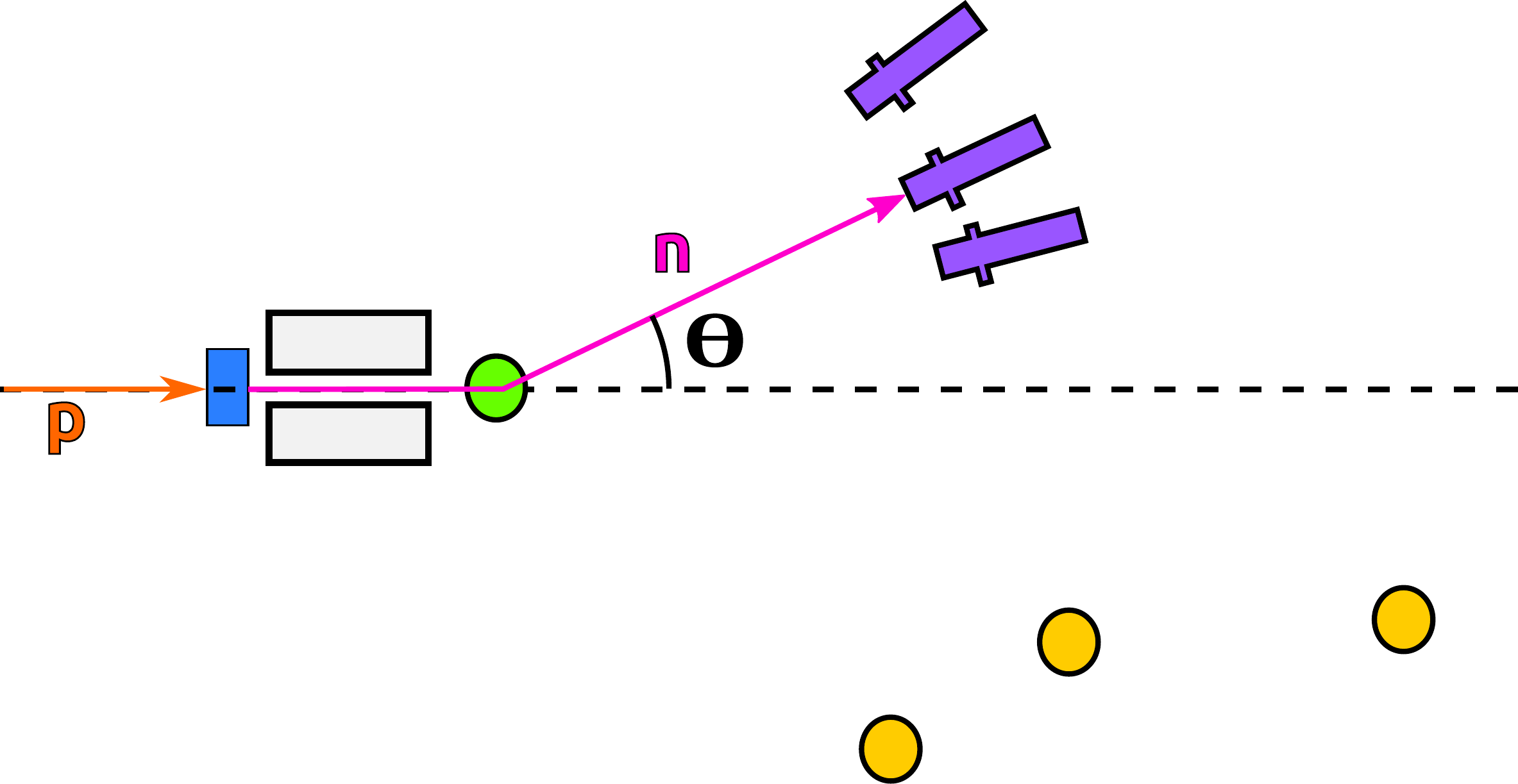}
 \caption{Drawing of the experimental setup. A proton (\emph{orange}) hits a LiF target (\emph{blue}) to produce a neutron (\emph{magenta}), which passes through a polyethylene collimator (\emph{white}) and scatters off of the liquid scintillator target (\emph{green}) at an angle $\theta$. The neutron then scatters in one of the 2'' coincidence detectors (\emph{purple}) or one of the 5'' coincidence detectors (\emph{yellow}), depending on $\theta$. 
 }
 \label{fig:nd_setup_dwg}
\end{figure}

The measurements described in this document were performed at the FN tandem accelerator at the University of Notre Dame's Nuclear Science Laboratory. The experimental setup is depicted in Figure~\ref{fig:nd_setup_dwg}. For these studies, a bunched proton beam from the accelerator was incident upon a LiF target, which produced neutrons through the \lith$(p,n)$\ber\ reaction at a mean energy of 690\,keV at 0$^\circ$ scattering angle. Calculations of the outgoing neutron energy-angle distribution are discussed in more detail in~\cite{westerdale_thesis}. More details about the proton beam and the LiF target are discussed in~\cite{xu_scintillation_2015}. Forward-scattering neutrons were selected by a polyethylene collimator (22\,cm$\phi\times$22\,cm, with a 2.5\,cm$\phi$ bore hole).

The primary liquid scintillator detector was placed at the end of the collimator, 50\,cm away from the LiF target. Three  2'' and three 5'' coincidence detectors were arranged at various angles relative to the proton beam. Time-of-flight cuts between the primary liquid scintillator detector, the coincidence detectors, and the beam pulser allowed us to identify particles and obtain clean sets of neutron recoils.

By selecting events in which a signal was seen in the primary detector and in one of the coincidence detectors, we could determine the scattering angle of the neutron. From this scattering angle, we determined the energy of the recoil, which we compared to the number of photoelectrons (\pe) detected. The angles from the center of the liquid scintillator detector to the centers of each of the coincident detectors relative to the proton beamline, along with the mean energy corresponding to each recoil angle, as determined by \geant~\cite{Agostinelli:2003fg} simulation, are summarized in Table~\ref{tab:detector_setup}. The uncertainties shown in the scattering angles are from the uncertainties in the positions of the detectors, which we estimate to be $\sim1$\,cm or $\sim2$\,cm, in each direction for the 2'' and 5'' detectors, respectively.

\subsection{Primary liquid scintillator detector}	

\begin{figure}[htb]
 \centering
 \includegraphics[width=0.15\linewidth]{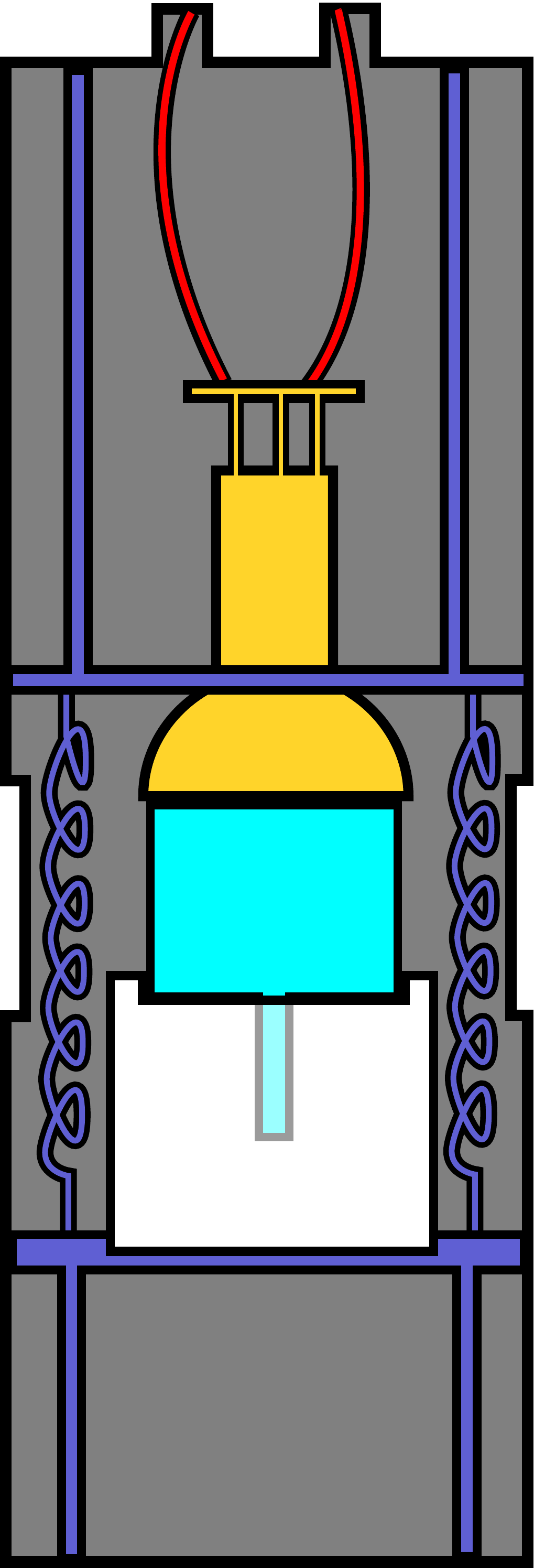}
 \caption{Drawing of the canister holding the scintillator cell. (\emph{Cyan}) The scintillator cell. (\emph{Yellow}) The \pmt. (\emph{Gray}) The stainless steel canister. (\emph{red}) The high voltage and signal cables. (\emph{White}) The PTFE support cup. (\emph{Purple}) Support structures and springs holding everything in place.}
 \label{fig:notredame_candwg}
\end{figure}

The primary liquid scintillator detector is a stainless steel cannister holding a 7.6\,cm$\phi\times$7.6\,cm fused silica cell, which contains the scintillator. The walls of the fused silica cell are $\sim$1--2\,mm thick. A 7.6\,cm$\phi$ Hamamatsu R11065 photomultiplier tube (\pmt) is optically coupled to the cell with optical coupling gel. A diagram of this setup is shown in Figure~\ref{fig:notredame_candwg}.

The sides, bottom, and stem of the cell are covered in a layer of Lumirror E6SR (188\,$\mu$m thick) reflector to increase the detector's light yield. The cell is supported by a PTFE cup. A steel support structure consisting of rods and springs holds the cell and \pmt\ in place; the stainless steel canister was thinned to $\sim1.4$\,mm around where the cell sits to reduce the amount of material that neutrons must pass through before reaching the scintillator.

The primary scintillator in the fused silica cell is \pc\ (C$_6$H$_3$(CH$_3$)$_3$). 
Boron is added by mixing the \pc\ with an equal volume of \tmb\ (BO$_3$(CH$_3$)$_3$). 
A total of 3\,g/L of \ppo\ is dissolved in this mixture to shift the \pc\ scintillation light into the visible spectrum.

\subsubsection{Cell preparation}

It is important that the scintillator be as pure as possible to avoid contaminants suppressing the scintillation light; it is especially important to avoid oxygen and water impurities, which may react with the \pc\ and \tmb.

To obtain a high-purity scintillator, we separately distilled \pc\ and \tmb\ in a dry nitrogen environment, following the procedure outlined in more detail in~\cite{westerdale_prototype_2016,westerdale_thesis}. The distillation process involved boiling each liquid in a three-neck flask heated by a heating mantle, while a stirring rod and boiling stones facilitated and steadied the boiling. Vapors traveled down a condenser, and the resulting fluid was collected in a flask. A steady nitrogen flow maintained a dry, nitrogen-rich environment. 

Distilled fluids were then transferred to a nitrogen-filled glove box, where they were measured and mixed together with the \ppo. The mixed scintillator was then transferred to a fused silica cell with a narrow neck. 

To further remove oxygen impurities from the scintillator, we transferred the cell to a manifold, where we bubbled nitrogen into the scintillator for several minutes. The cell was then frozen by submerging it into a liquid nitrogen bath, and a vacuum was pulled on the cell while a torch was used to seal the cell at the neck.

\subsection{Coincidence detectors}
Two different types of detectors are used as coincidence detectors. The first is a 5.1\,cm$\phi\times$5.1\,cm Eljen 510-20$\times$20-9/301 detector, and the second is a 12.7\,cm$\phi\times$12.7\,cm Eljen 510-50$\times$50-1/301 detector. Both models use the commercial EJ-510 reflector and the EJ-301 liquid scintillator, which is a xylene-based scintillator with wavelength-shifting organic fluors. 
These detectors have a typical light yield of $\sim1$\,PE/\keVee.

\subsection{Electronics and data acquisition}\label{subsec:daq}

\begin{figure}
 \centering
 \includegraphics[width=0.6\linewidth]{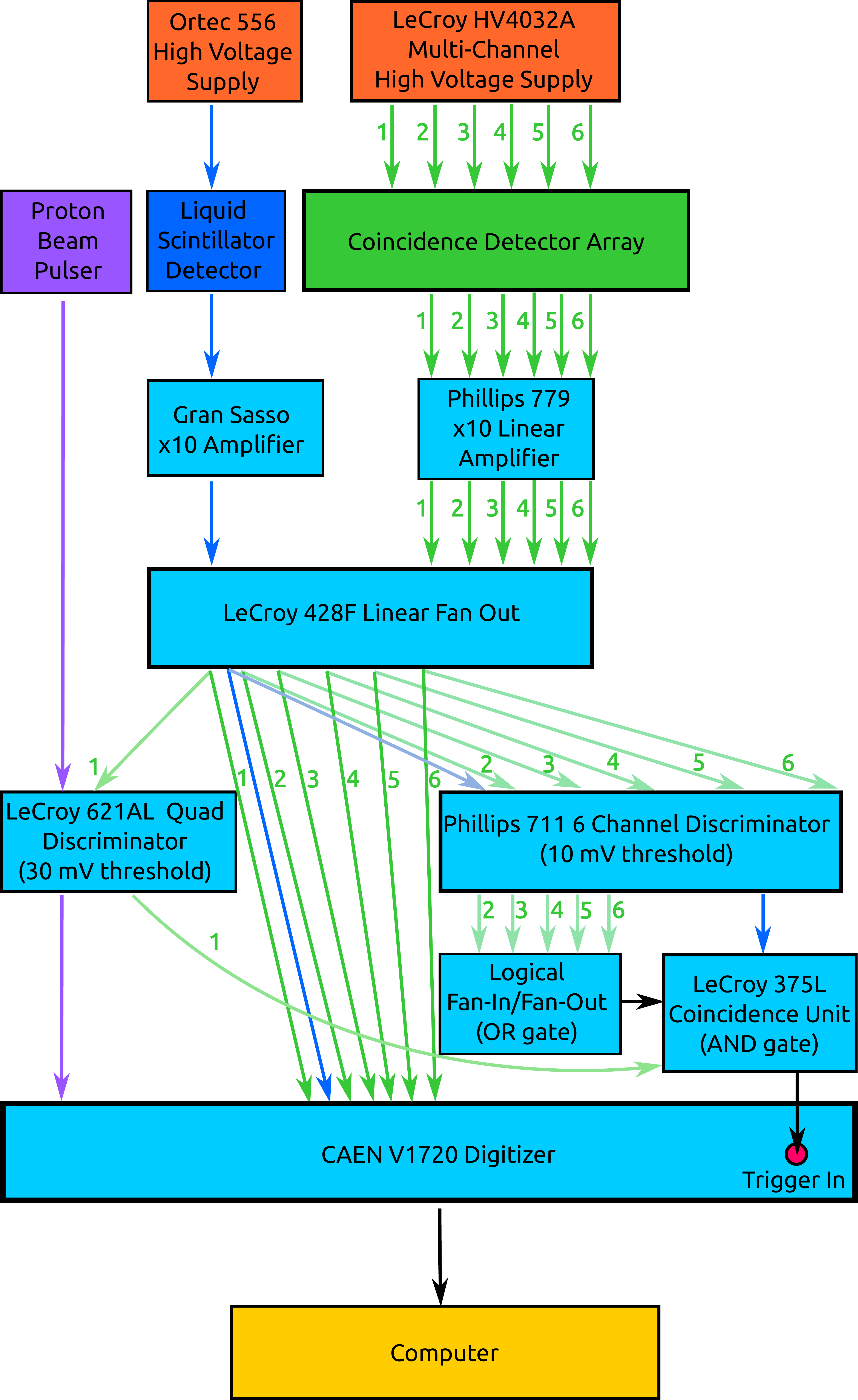}
 \caption{A schedmatic diagram of the electronics and data acquisition system.}
 \label{fig:nd_daq}
\end{figure}

A schematic diagram describing the electronics and data acquisition (DAQ) system is shown in Figure~\ref{fig:nd_daq}. 

The \pmt\ in the liquid scintillator detector was powered by an Ortec 556 high voltage power supply operating at 1.4\,kV and 0.4\,mA. The coincidence detector \pmt s were powered by a LeCroy HV4032A multi-channel high voltage supply at 1.15--1.6\,kV.

The DAQ was setup to trigger on coincidences between the liquid scintillator detector and one of the six coincidence detectors. Additionally, a periodic signal produced by the proton beam pulser gave timing information that differs from the time at which a neutron is produced in the LiF target by an approximately constant offset. This periodic signal was used to make time-of-flight cuts. These cuts are useful for particle identification by differentiating between \grs\ and neutrons by their speed.

Signals produced by the pulser were read by a LeCroy 621AL quad discriminator set to a 423\,mV threshold. Logic pulses produced by this discriminator were read by a CAEN V1720 digitizer (12 bit, 250\,MS/s). 

Signals from the liquid scintillator detector were amplified by a $\times10$ front-end amplifier module developed at the Laboratori Nazionali del Gran Sasso, while signals from the coincidence detectors were amplified by a Phillips 779 $\times10$ amplifier. These amplified signals were passed to a LeCroy 428F linear fan out module that sent one copy of each signal to the CAEN V1720 digitizer. Another copy of the signal from the liquid scintillator detector was sent to the LeCroy 621AL quad discriminator, and a second copy of the signals from the coincidence detectors was sent to a Phillips 7116 discriminator with a 10\,mV threshold. This threshold was set to the amplitude of approximately 1.5 \pe\ read out from each detector.

Logic pulses produced by the Phillips discriminator were passed to a fan-in/fan-out module that was used as a logical OR gate, which produced a logic pulse whenever at least one coincidence detector produced a signal above the discriminator threshold. This pulse was passed to a LeCroy 375L coincidence unit, which was used as an AND gate with a 400\,ns coincidence window. The pulse from the LeCroy 621AL discriminator was also passed to this module, and the AND gate produced a logic pulse when it received a pulse from both the OR gate and the liquid scintillator detector discriminator.

The signal from the AND gate was passed to the trigger of the CAEN V1720 digitizer. When this trigger signal was received by the digitizer, it recorded the waveforms from the proton beam pulser, liquid scintillator detector, and the coincidence detectors to disk. The recorded waveform started 2\,$\mu$s before the trigger signal was received and lasted for a total of 20\,$\mu$s. 

\begin{table}[htb]
 \centering
 \caption{The diameter, scattering angle, distance from the liquid scintillator detector, and mean recoil energy in the liquid scintillator detector corresponding to events with coincidences in each of the coincidence detectors. The uncertainty in the recoil energy is dominated by the uncertainty in the scattering angle.}
 \begin{tabular}{cccccc}\hline\hline
  Detector & Diameter & $\theta$                & Distance      & $E_R$                 \\
           & [cm]     & [deg]                   & [cm]          & [keV]                 \\\hline\rule{0pt}{3ex}
  1        & 12.7     & 59.02$^{+1.02}_{-1.03}$ & 150.3$\pm$2.4 & 467.2$^{+7.9}_{-8.0}$ \\\rule{0pt}{2.5ex}
  2        & 12.7     & 41.26$^{+1.08}_{-1.08}$ & 148.8$\pm$2.4 & 274.9$^{+9.1}_{-9.1}$ \\\rule{0pt}{2.5ex}
  3        & 12.7     & 24.88$^{+0.76}_{-0.75}$ & 200.0$\pm$2.4 & 106.1$^{+5.3}_{-5.2}$ \\\rule{0pt}{2.5ex}
  4        & 5.1      & 48.04$^{+1.16}_{-1.15}$ & 69.9$\pm$2.0  & 355.4$^{+9.8}_{-9.8}$ \\\rule{0pt}{2.5ex}
  5        & 5.1      & 32.37$^{+1.13}_{-1.14}$ & 69.6$\pm$2.0  & 172.7$^{+9.0}_{-8.9}$ \\\rule{0pt}{2.5ex}
  6        & 5.1      & 18.13$^{+1.03}_{-1.05}$ & 69.7$\pm$2.0  & 56.9$^{+6.0}_{-5.9}$  \\\hline\hline
 \end{tabular}
 \label{tab:detector_setup}
\end{table}

\section{Data analysis}\label{sec:analysis}
As discussed in Section~\ref{sec:setup}, data were taken at 6 different neutron scattering angles, each corresponding to a different proton recoil energy in the primary liquid scintillator detector. Neutrons were selected in the detectors based on their time-of-flight relative to the pulser; coincident \grs\ travel faster and therefore appear sooner than neutrons in the detectors. The mean proton recoil energies measured are summarized in Table~\ref{tab:detector_setup}, and cover nuclear recoil energies in the range of 56.9--467.2\,keV.

While carbon, oxygen, and boron recoils were also present in the data, the higher masses of the nuclei caused the recoil signals to be quenched below the detector's sensitivity. However, since neutrons will lose most of their energy to proton recoils as they slow down, describing proton recoil quenching is sufficient for understanding the detector's response to neutron scatters. We therefore do not report the quenching factors of the other nuclei here.

Data were taken over the course of two days, for a total beam time of $\sim$28 hours. Additional runs were taken with an \amer\ source before and after the beam runs for light yield calibration. \amer\ \alp-decays and produces a 59.54\,keV \gr. Runs with this sources provided light yield calibration and allowed us to check for changes in the detector light yield throughout the course of the runs. 

In order to determine the light yield of the liquid scintillator detector, we placed a \amer\ source just outside the detector.
An \amer\ source was chosen for the light yield calibration because the \gr\ produced by the source is low enough energy to produce a clean full energy peak in the detector, with little probability of the \gr\ escaping without depositing all of its energy. Additionally, since we expect this source to produce a comparable number of photoelectrons to the proton recoils we are considering, we were able to preserve the high gain settings in the electronics without saturating the digitizers. Using this source for calibration assumes that the scintillator response remains linear for electron recoils between $\sim$60\,keV and higher energies.
The \gr\ spectrum in the detector was recorded, and a Gaussian peak corresponding to the full-energy absorption of the \gr\ was observed. Fitting a Gaussian distribution to this peak gave a mean of 81.47$\pm$0.10\,\pe, corresponding to a light yield of 1.368$\pm$0.002\,\pe/keV. The light yield was measured before and after these measurements, and no significant variation was observed. 

The mean single photoelectron charge in each phototube was monitored through the course of each run, as was the collected charge distribution of each event. No significant change in either quantity was seen over time.

A method of trigger efficiency calibration is described in~\cite{xu_scintillation_2015}. Due to the fast scintillation decay time of this scintillator, with a dominant component around $\sim3$\,ns~\cite{alimonti_large-scale_1998,okeeffe_scintillation_2011}, we found that the loss in trigger efficiency above $\sim4$\,\pe\ is negligible. We therefore did not include any corrections for the trigger efficiency in this analysis.

\subsection{Data processing}
The data recorded to disk, as described in Section~\ref{subsec:daq}, consists of waveforms for each of the six coincidence detectors and the primarily liquid scintillator detector, measured in ADC counts, and a periodic waveform from the proton beam pulser. The software and algorithms used to process the data are described in more detail in~\cite{xu_scintillation_2015}; a summary is given here.

The baselines of the waveforms were subtracted using a drifting baseline-finding algorithm that suppresses low-frequency electronic noise without removing scintillation pulses. Waveforms in which a clear baseline could not be identified or the digitizers were satured were discarded. Single photoelectrons were identified from the tails of scintillation pulses, and the mean single photoelectron charge was determined by fitting a Gaussian plus an exponential to the single photoelectron charge distribution. The single photoelectron charge was then used to normalize each waveform so that integrals over the waveform can be measured in \pe. 

All pulses with an amplitude greater than $\sim0.2$\,\pe\ were tagged, and the integrals of those in the first 200\,ns were grouped together in order to include the total charge produced in each scintillation event.

\begin{figure}[htb]
 \centering
 \includegraphics[width=\linewidth]{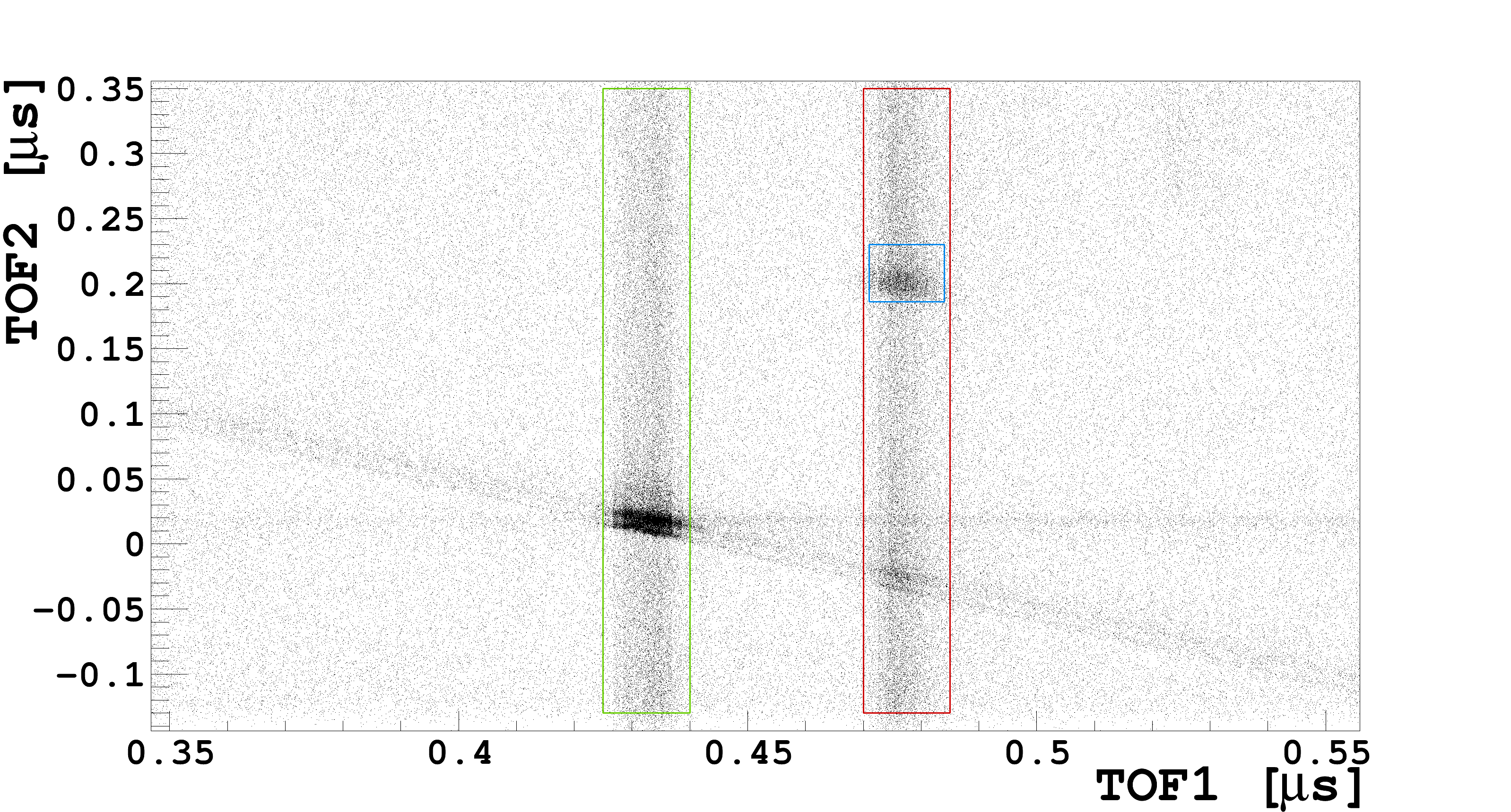}
 \caption{The time-of-flight between the pulser and the liquid scintillator detector (TOF1) versus the time-of-flight between the liquid scintillator detector and coincidence detector number 3 (TOF2). The green box on the left shows events where a \gr\ from the LiF target scattered in the liquid scintillator detector and a random coincidence background triggered the coincidence detector. The red box on the right shows events where a neutron scattered in the liquid scintillator detector. Events in the blue box then had the neutron scatter in the coincidence detector, while those in the red box had a random background trigger it.}
 \label{fig:nd3_tof}
\end{figure}

Time-of-flight cuts were used to identify events in which a neutron scattered in the liquid scintillator detector and then in one of the coincidence detectors. Two time-of-flight variables were defined for each coincidence: TOF1 is the time between the proton beam pulser and the signal in the liquid scintillator detector, which corresponds to the time it takes a particle to travel from the LiF to the detector up to a constant offset, and TOF2 is the time between the signal in the liquid scintillator detector and the triggered coincidence detector. 

Figure~\ref{fig:nd3_tof} shows how these two variables relate to each other for events with a coincidence in detector 3, which corresponds to a mean proton recoil energy of 106.1\,keV. The vertical band on the left of this plot corresponds to events where a \gr\ from the LiF target scattered in the liquid scintillator detector and coincidence detector 3 had a random background. Events in the band on the right correspond to cases where a neutron from the LiF scattered in the liquid scintillator detector; those in the cluster of points around TOF2$\approx0.2$\,$\mu$s correspond to cases where the neutron then scattered in the liquid scintillator detector, while the rest correspond to cases where a random coincidence was seen in this detector or the neutron scattered multiple times before reaching the coincidence detector and therefore had a different time of flight. 

We attribute events in the diagonal band (where TOF1+TOF2 is constant) to cases where a \gr\ left the collimator at an angle and hit the coincidence detector directly while a random background triggered the liquid scintillator detector. The gap between the two parallel bands within this band is likely due to the 4\,ns sample width and the finite timing resolution of the DAQ. 

We attribute events in the band with TOF2 between 0 and 0.05\,$\mu$s to random background \grs\ scattering in one detector and the other. Such \grs\ may come from environmental radiation or from earlier neutrons that have thermalized in the liquid scintillator and produce a 478\,keV \gr\ after capturing on \borten.

Neutrons that scatter multiple times in the liquid scintillator detector may provide a background, since their recoil energy and angle may not be correlated when they leave the detector. However, neutrons that scatter multiple times before leaving in a certain direction will have lower kinetic energy than those that only scattered once, and so a narrow TOF2 cut can be used to reduce the effects of multiple scattering. 

%
%

\subsection{Quenching factor evaluations}
When neutrons scatter in the scintillator, they predominantly lose energy by scattering off of hydrogen. We therefore measure the signals produced by various proton recoil energies due to these neutron scatters. In order to evaluate the quenching factor of proton recoils at each of the sampled energies, we performed a \geant\ simulation of this experiment, replicating the cuts in the data in the simulation. We then produced an energy deposition spectrum in the liquid scintillator detector for events with coincidences in each of the six coincidence detectors. The advantage of these simulations over kinematic calculations is that they include whatever effects of multiple scattering persist through the TOF2 cut, as well as effects from the spread in recoil energy due to the finite sizes of the detectors, and interactions the neutrons may have with other present materials.

In order to determine what quenching factors are consistent with the data, we convolved each of these simulated spectra with a Gaussian with mean $\mu$ and variance $\sigma^2$ given by
\begin{eqnarray*}
 \mu      &=& E\times LY_{rel} \\
 \sigma^2 &=& (1+\sigma^2_{\text{SPE}})\times E\times LY_{rel}
\end{eqnarray*}
where $E$ is the energy deposited, $LY_{rel}$ is the relative light yield of protons at this recoil energy, and $\sigma^2_{\text{SPE}}$ is the relative variance of the single photoelectron charge, determined to be $\sim10\%$ from the single photoelectron charge calibration.

The variance of the response function can be broken into two terms: $E\times LY_{rel}$ and $\sigma^2_{\text{SPE}}\times E\times LY_{rel}$. The first term is equal to the mean number of photoelectrons detected at a given energy and represents the Poisson variance. 
The second term represents the variance of the charge measured from these photoelectrons.

\begin{figure}[htb]
 \centering
 \includegraphics[width=0.7\linewidth]{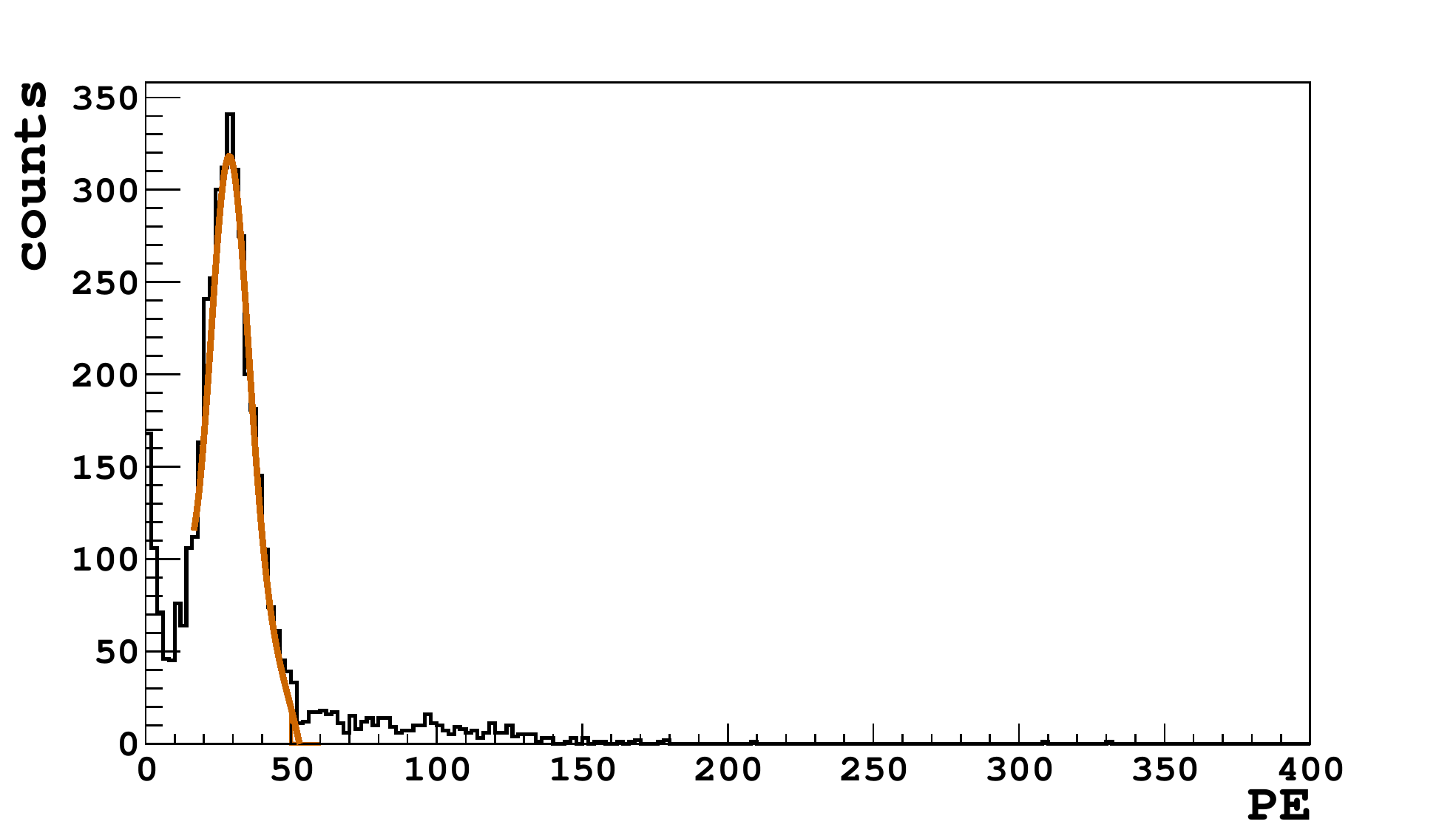}
 \caption{(\emph{Black}) The measured energy spectrum of events with neutrons scattering in the liquid scintillator detector and then in coincidence detector 2, at 41.26$^\circ$ from the beamline, corresponding to a mean proton recoil energy of 274.9\,keV. (\emph{Red}) The simulated spectrum fit to the data around the peak at 28.3\,\pe.}
 \label{fig:nd_coin2_spec}
\end{figure}

Allowing $LY_{rel}$ and an overall rate constant to be free parameters, we fit the simulated spectra to the data (convolved with a Gaussian) and determined the quenching factor at each proton recoil energy by dividing $LY_{rel}$ by the measured detector electron recoil light yield. The quenching factor $QF$ is therefore given by $QF=LY_{rel}/LY$, where $LY$ is the light yield measured by the \amer\ source. Figure~\ref{fig:nd_coin2_spec} shows this fit performed for events in which coincidence detector 2 was triggered, corresponding to a mean proton recoil energy of 274.9\,keV. Since quenching introduces nonlinearity to the scintillator response, the fit was limited to the area around the peak---in this case, the fit was performed in the range 10--50\,\pe. 


\begin{table}[htb]
 \centering
 \caption{The expected mean recoil energy for neutron scatters with coincidences in each detector, the observed peak location, and the corresponding quenching factor (QF) at that proton recoil energy for the measurements reported here.}
 \begin{tabular}{cccc}\hline\hline
  Detector & $E_R$ [keV] & Peak [\pe] & QF    \\\hline
  1        & 467.2       & 73.4       & 0.115 $\pm0.002$ \\
  2        & 274.9       & 28.3       & 0.075 $\pm0.002$\\
  3        & 106.1       & 8.9        & 0.062 $\pm0.003$\\
  4        & 355.4       & 43.7       & 0.090 $^{+0.004}_{-0.002}$\\
  5        & 172.7       & 17.2       & 0.073 $\pm0.003$\\
  6        & 56.9        & 6.1        & 0.078 $\pm0.007$\\\hline\hline
 \end{tabular}
 \label{tab:fit_results}
\end{table}

\begin{figure}[htb]
 \centering
 \includegraphics[width=0.8\linewidth]{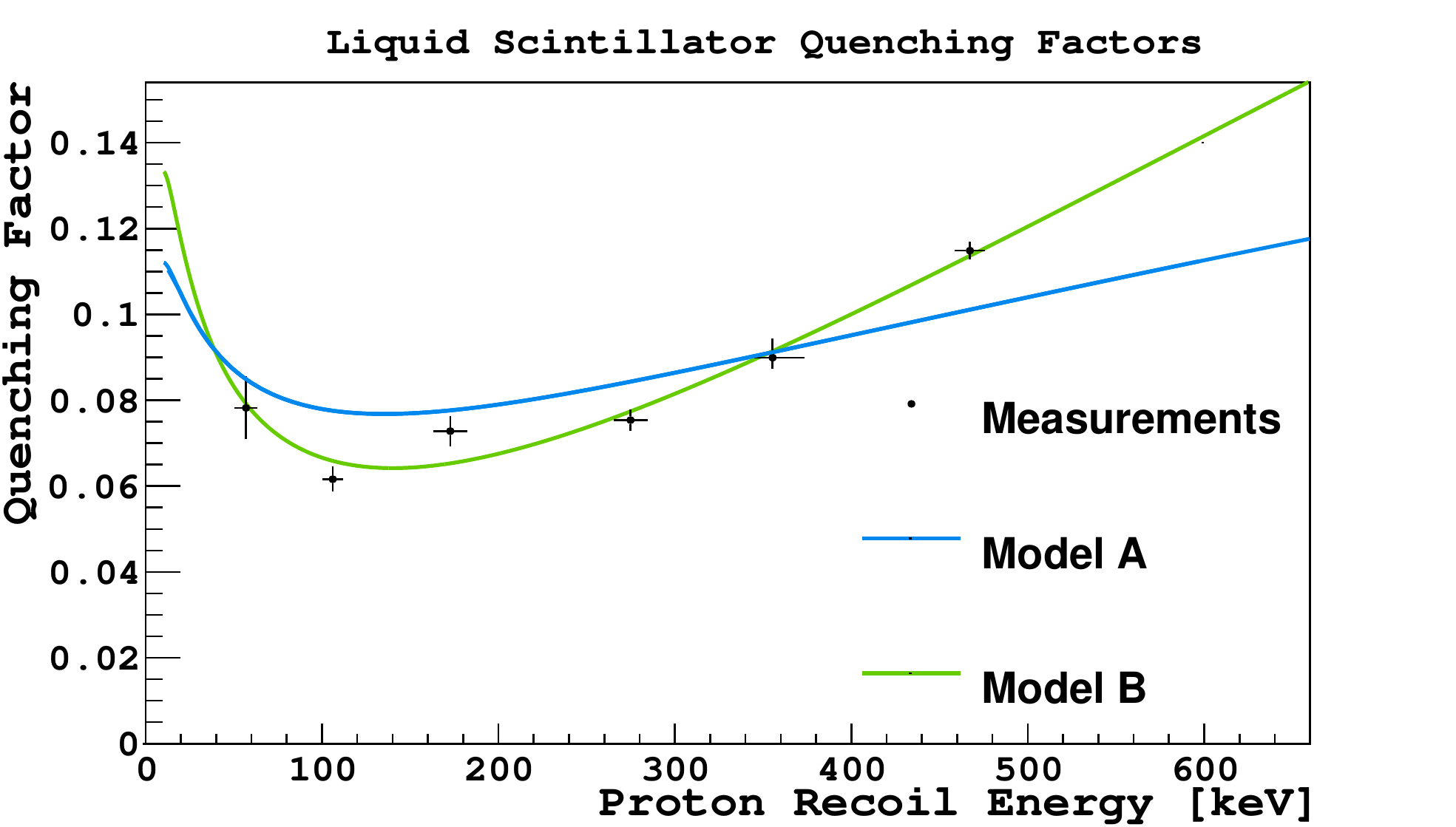}
 \caption{(\emph{Black}) Proton recoil quenching factors measured in this experiment. (\emph{Blue}) Model~A and (\emph{Green}) Model~B, fit to the data. Model~C lines up exactly with Model~A, and is therefore not visible here.}
 \label{fig:proton_qf_fit}
\end{figure}

The results of these fits are summarized in Table~\ref{tab:fit_results} and illustrated in Figure~\ref{fig:proton_qf_fit}.

The uncertainties in the quenching factor measurements are dominated by the uncertainty in the mean neutron scattering angle for coincidences with each detector, as shown in Table~\ref{tab:detector_setup}. 

\section{Models considered}

One of the most widely used models describing scintillation and quenching due to ionizing radiation is the one developed by Birks~\cite{birks_scintillations_1951}. Models that have since been developed largely build upon this framework.

The scintillation models considered here are given below. These models describe the amount of fluorescent photons produced per unit track length of the ionizing particle $dF/dx$ as a function of the stopping power $dE/dx$ of the ionizing radiation in the scintillator.

\begin{enumerate}[A)]
 \item $\frac{dF}{dx} = \frac{S\cdot dE/dx}{1+kB\cdot dE/dx}$ \label{enum:birks}
 \item $\frac{dF}{dx} = \frac{S\cdot dE/dx}{1+kB\cdot dE/dx+C\cdot(dE/dx)^2}$ \label{enum:craun}
 \item $\frac{dF}{dx} = \frac{S_e\cdot\left(dE/dx\right)_e+S_n\cdot\left(dE/dx\right)_n}{1+kB_e\cdot\left(dE/dx\right)_e+kB_n\cdot\left(dE/dx\right)_n}$ \label{enum:hong}
\end{enumerate}

Model~\ref{enum:birks} is Birks' model, where $kB$ is the empirically determined Birks constant used to parameterize the model as a function of the total stopping power $dE/dx$. The constant term $S$ is the scintillation yield of the scintillator, denoting the number of photons produced per unit energy deposited. When we calculate quenching factors and divide by the light yield of the scintillator, this $S$ term is divided out, since it is the same for electron and nuclear recoils. The only fit parameter in this model is $kB$. This model is discussed in~\cite{birch_evaluation_2015} as a description of the creation of ``damaged'' molecules and ions that may be produced in the track of the ionizing particle; these species may non-radiatively dissipate energy, decreasing the amount of scintillation light produced.

Model~\ref{enum:craun} is an extension to Birks' model discussed in~\cite{craun_analysis_1970} that adds a term proportional to $(dE/dx)^2$ by the empirically determined constant $C$. Since $S$ divides out when we normalize by the detector light yield, this model has two fit parameters: $kB$ and $C$. This model was first theoretically proposed in~\cite{chou_nature_1952}, and postulates that scintillation quenching has a term quadratic in the stopping power. Such a term may arise due to bi-excitonic interactions, as discussed in~\cite{voltz_radioluminescence_1968} in the context of scintillation pulse shape, that allow two excited \pc\ molecules to non-radiatively de-excite. For example, two \pc\ molecules excited into the triplet state $T_1$ may undergo the reaction $T_1+T_1\rightarrow T_1 + S_0$ to produce a ground state molecule $S_0$ at a rate proportional to the square of the triplet density.

Model~\ref{enum:hong} is discussed in~\cite{hong_scintillation_2002} and allows energy lost to electron and nuclear recoils to scintillate and be quenched separately. This model therefore distinguishes between the electronic stopping power $dE/dx|_e$ and the nuclear stopping power $dE/dx|_n$; $S_e$ and $S_n$ are the scintillation yields for energy lost to electrons and nuclei, respectively, due to these different mechanisms. Similarly, $kB_e$ and $kB_n$ characterize quenching for electron and nuclear recoil energy loss terms. When we calculate quenching factors by dividing by the electron recoil light yield, we are left with three fit parameters: $S_n/S_e$, $kB_e$, and $kB_n$.

\section{Model fits and results}
In order to compare these models to the measured quenching factors, we used SRIM~\cite{ziegler2010srim} to calculate the stopping powers in the scintillator and integrated over the track of the recoiling proton nucleus as it slows down. Ziegler et al. report that proton stopping powers computed by SRIM, which show a Bragg peak around $\sim70$\,keV, tend to agree with measurements within 3.9\%. This Bragg peak explains the decrease in quenching below $\sim100$\,keV, as shown in Figure~\ref{fig:proton_qf_fit}.

Figure~\ref{fig:proton_qf_fit} shows the data presented in Table~\ref{tab:fit_results}. We show fit results for the three models to the data. The results of these fits are summarized in Table~\ref{tab:model_fits}. We found that Model~\ref{enum:hong} converged to Model~\ref{enum:birks} in this fit because $dE/dx|_e$ dominates $dE/dx|_n$ for protons in the scintillator. Therefore, distinguishing between electron and nuclear contributions to the stopping power cannot explain the differences between Model~\ref{enum:birks} and the data.

\begin{table}
 \centering
 \caption{Summary of fits of scintillation quenching Models~A,~B, and~C explored here.}
 \begin{tabular}{l|cccc}\hline\hline
        & $kB$\,[cm/MeV]    & $C$\,[cm$^2$/MeV$^2$]&           & $\chi^2/$NDF \\\hline
  A     & $0.0153\pm0.0002$ & ---                  &           & 19.4         \\ 
  B     & 0                 & $(2.19\pm0.04)\E{-5}$&           & 1.6          \\\hline\hline
        & $kB_e$\,[cm/MeV]  & $kB_n$\,[cm/MeV]     & $S_n/S_e$ & \chisqndf    \\\hline
  C     & $0.0153\pm0.0002$ & $0.06\pm0.06$        & $0\pm0.8$ & 32.3         \\\hline\hline
 \end{tabular}

 \label{tab:model_fits}
\end{table}

As shown in Figure~\ref{fig:proton_qf_fit} and Table~\ref{tab:model_fits}, Model~\ref{enum:craun} describes the data much better than the other two models do. Furthermore, we have found that the best fit of Model~\ref{enum:craun} has $kB = 0$, showing that the term quadratic in the stopping power dominates the quenching factor. This model fit the data with \chisqndf = 1.6, significantly better than the \chisqndf = 19.4 fit we obtained using Model~\ref{enum:birks}.

The quadratic dependence on the stopping power in Model~\ref{enum:craun} causes the model to predict greater variation in the stopping power with the proton recoil energy, allowing it to better describe our measurements. Since Models~\ref{enum:birks} and~\ref{enum:hong} vary linearly with the stopping power, they were not able to adequately describe the variation we saw, despite the greater number of fit parameters in Model~\ref{enum:hong}. 
We therefore conclude that the success of Model~\ref{enum:craun} over the other two models is due to the quadratic dependence on the stopping power, which may result from interactions between excitons in the scintillator that allow them to non-radiatively de-excite.

\section{Conclusions}
We have measured proton recoil quenching factors at six different energies ranging from  56.9--467.2\,keV for a boron-loaded organic liquid scintillator, composed of an equal volume of \pc\ and \tmb, with 3\,g/L of \ppo. 
We found that Model~\ref{enum:craun} best describes the data.
This agreement supports the theory of scintillation quenching put forth in~\cite{chou_nature_1952}. We find that a term quadratic in $dE/dx$, as might arise due to bi-excitonic quenching processes, is needed to describe the quenching of the scintillator. 

These measurements and this model will allow experiments that use such scintillators to accurately reconstruct the energy of nuclear recoils. This ability is important for fast neutron spectrometry, where proton recoil energy can be used to determine neutron energies, and for thermal neutron detection, which relies on detecting the nuclear recoil energy deposited by the \borten$(n,\alpha)$\lith\ reaction products. Furthermore, this model can be instrumental to simulating boron-loaded neutron detectors to ensure that a design will have the desired sensitivity.

While these results will be instrumental in modeling the response of the \dsf\ neutron veto, which uses a similar cocktail, they may also help model the gadolinium-loaded scintillator that will be used in the neutron veto of the LZ dark matter experiment~\cite{malling_after_2011}.

\acknowledgments
We thank University of Notre Dame for hosting this experiment and lending us equipment. We thank Stephen Pordes for lending us neutron detectors and equipment, Mike Souza helping make the scintillator cell, and Ben Loer for developing the DAQMAN software used for this study. This study was supported by NSF Grants No. PHY-1242625 and No. PHY-1419765. F. Froborg was supported by the Swiss National Science Foundation.

\bibliographystyle{JHEP}
\bibliography{biblio}

\end{document}